\documentclass[notitlepage]{article}

\usepackage{a4wide}
\usepackage{graphicx}

  \renewenvironment{thebibliography}[1]{%
    \begin{oldthebibliography}{#1}
      \setlength{\itemsep}{0.9ex}%
  }%
  {%
    \end{oldthebibliography}%
  }

\begin{document}
\renewcommand{\refname}{Publications}

\begin{titlepage}

\newlength{\Size}
\setlength{\Size}{0.2\textwidth}
\newlength{\Shift}
\settoheight{\Shift}{L}
\addtolength{\Shift}{-\Size}


\begin{center}
\bf \Huge Leslie Hodson, particle physicist and pioneer in the study of cosmic rays\\
\Large 1925 -- 2010\\
\vspace{0.5cm}

{\includegraphics[width=9cm]{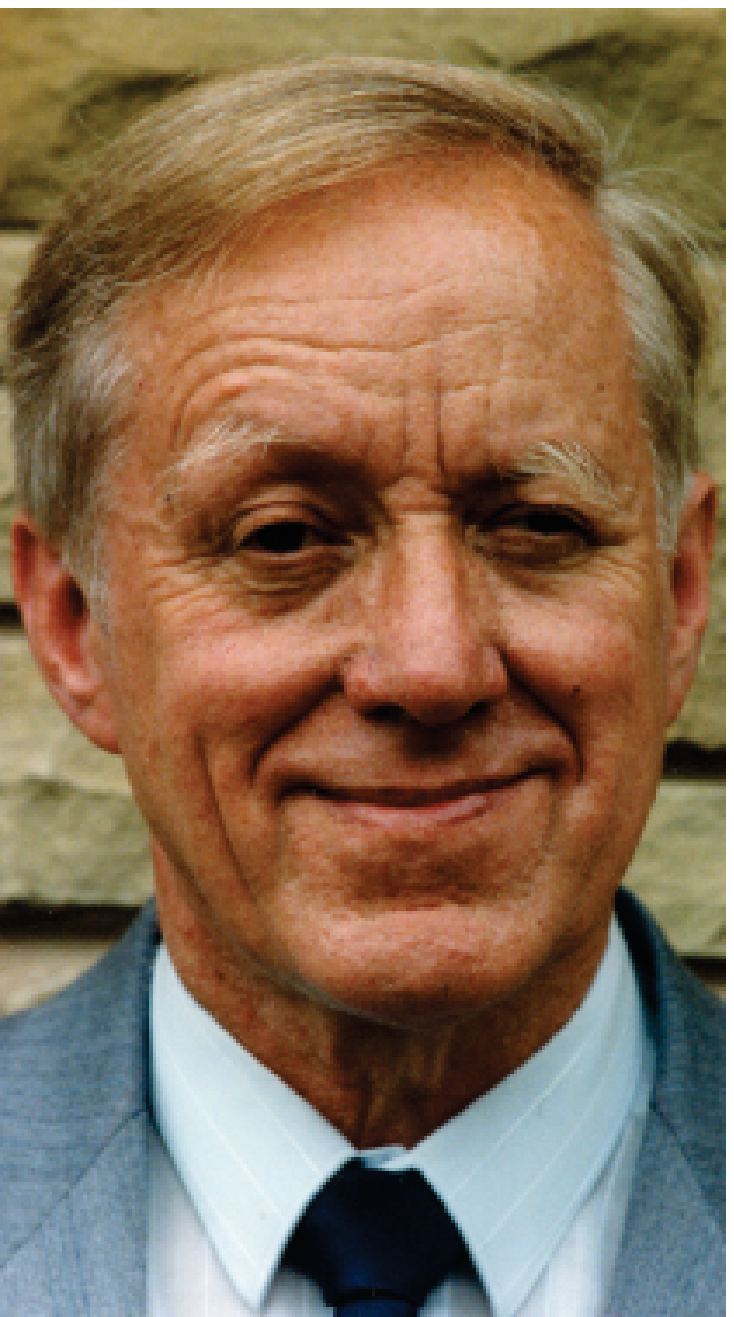}}
\\
\vspace{0.5cm}

\Large{Obituary and Bibliography}
\end{center}

\normalsize\rm 
 

\end{titlepage}

\suppressfloats[t]   

\section*{Leslie Hodson}

Leslie Hodson was an experimental particle physicist who developed
cloud and spark chamber techniques in the study of cosmic rays.
Yet, despite his achievements in the field, he believed that the most
useful thing he did was prevent asbestos from being used in the
construction of the new campus at the University of Leeds in the 1960s.

Albert Leslie Hodson was born an only child in 1925 to tenant farmer
parents at Fishlake, on the floodplain of the River Don, north-east 
of Doncaster. He was an only child, but had an extended family of 
aunts, uncles and cousins, all living in the near vicinity.  
He first attended
Fishlake Endowed School, a two-room school where children of all
levels were taught together.  He failed his 11-plus but despite this,
his father found him a place at Thorne grammar school, at a cost of
three guineas a term, and his parents managed to scrimp enough to pay
for it.

In 1943, Leslie's higher school certificate results were good enough
for a county scholarship.  He decided to study physics at Manchester
University, having heard from a cousin that Professor P.M. Blackett
was working there.  Leslie had never heard of Blackett, but what had
impressed his cousin was good enough for him.  After two years he was
recruited to the Royal Aircraft Establishment at Farnborough, but the
week he was due to report he developed mumps.  The war ended, and
instead of going to Farnborough he completed his third year at
Manchester.

During this year he devised his own project on the optics of sodium
vapour using equipment of his own construction.  Blackett immediately
offered him a PhD researching cosmic rays.

He studied altitude effects in cosmic ray air-showers.  This involved
building his own instruments, including making Geiger counters from
glass tubing and fitting them into bomb casing.  This instrument
package was loaded into the bomb bay of a Mosquito and flown to over
30,000ft (9.1km).  The solution to a problem in the use of Geiger tubes led to
his first scientific paper at only 23.

The mainstay of Blackett's cosmic ray research was the cloud chamber.
He had received the Nobel Prize in 1948 for his work on triggering
chambers using external Geiger counters and the subsequent discovery
of the positron.  In a flash of inspiration, Leslie realised that the
trigger could be generated from the gas within the chamber itself.
Despite others' doubts, he persuaded Blackett to support him.  He also
built the electronics to automatically control the chamber, secretly,
out of hours, protecting it from BlackettÕs critical view by a cloth
cover and only revealing its existence when it was fully working.  As
a result of all this work, Manchester
appointed him assistant lecturer.

In 1951 he became a research associate at Princeton.  The Princeton
cosmic ray group was running cloud chambers at Echo Lake in the
Rockies at an altitude of 3,230m.  They were studying the so-called
V-particles, recently discovered at Manchester.  All the images were
photographed and, on the last reel of film they discovered a new
particle, now known as the K+, one of the kaons.  Their measurement of
the particle's rest mass was within two per cent of the currently
accepted figure.

Leslie returned to Britain in 1954 to take up a lectureship at the
University of Leeds.  He designed and built the world's largest cloud chamber,
having taught himself the required skills, and made deft use of local
industry in its construction.

The mid-Sixties were a period of expansion in the universities, and
Leeds had chosen Chamberlin, Powell and Bon to design a new campus.
Leslie worked with them, providing a link between the architectural
vision and the scientists and educators who were to use the buildings.
The design called for asbestos cement panels as duct covers. Leslie
had read in the scientific press that asbestos was hazardous and
fought against its use.  The University banned the use of asbestos in
all future buildings, probably the first time this was done in
Britain.

By the end of the 1960s the physics department was installed in its
elegant modernist buildings and Leslie could turn once again to
research.  The quark model had been proposed by Murray Gell-Mann and
George Zweig in 1964 and physicists were looking for evidence of them.
Using the Leeds cloud chamber, he and his team searched for free
quarks in the cores of cosmic ray air-showers.  Quarks were predicted
to have 1/3 or 2/3 the charge of an electron and this
would give rise to 1/9 or 4/9 the ionisation along their
path.  This would be clearly visible by the thickness of the tracks
produced in a cloud chamber.  However, the cores of cosmic ray
air-showers have vast numbers of particle tracks, all of which had to
be examined in detail.  It was painstaking work and at the end of it
Leslie and his team came to the conclusion that there was no evidence
that free quarks existed.

The visual and photographic techniques Leslie used were both
labour-intensive and expensive and, towards the end of his career, this
became a seriously limiting factor.  More recent experiments replaced the
cameras with electronic techniques using charge read-out from vast
numbers of wires and strips.  However, now, in an era where
high-resolution digital photography is cheap and the computer power to
process it ubiquitous, a return to visual and photographic techniques
offers simplicity and elegance, and a further potential reduction in cost.

Leslie was a Methodist and a teetotaller.  He had a huge range of
interests, including philosophy, genealogy, planning, architecture,
music, education and modern art.  He was a founder member of his local
residents' association and fought inappropriate developments.  He was
also a keen gardener, winning prizes at his local horticultural
society for fruit and vegetables, including greenhouse-grown peaches.
\\

\noindent\emph{Albert Leslie Hodson, scientist: born Fishlake, West 
Riding of Yorkshire, 15 July 1925; married 1958 Joyce Wicks (three sons); died
Leeds, West Yorkshire, 1 March 2010.  }\\

\noindent A version of this obituary first appeared in ``The 
Independent'',
London,  27th May 2010.\\
I would like to thank Mugdha Joshi for considerable help in preparing 
the bibliography and the staff of the Independent for editorial 
assistance.\\
\\ 
\noindent John McMillan \\ 
\\
Department of Physics and Astronomy, \\
The University of Sheffield, Sheffield, South Yorkshire, S3 7RH, 
Great Britain.\\

\noindent j.e.mcmillan@sheffield.ac.uk\\
\pagebreak

\begin{center}
\includegraphics[width=15cm]{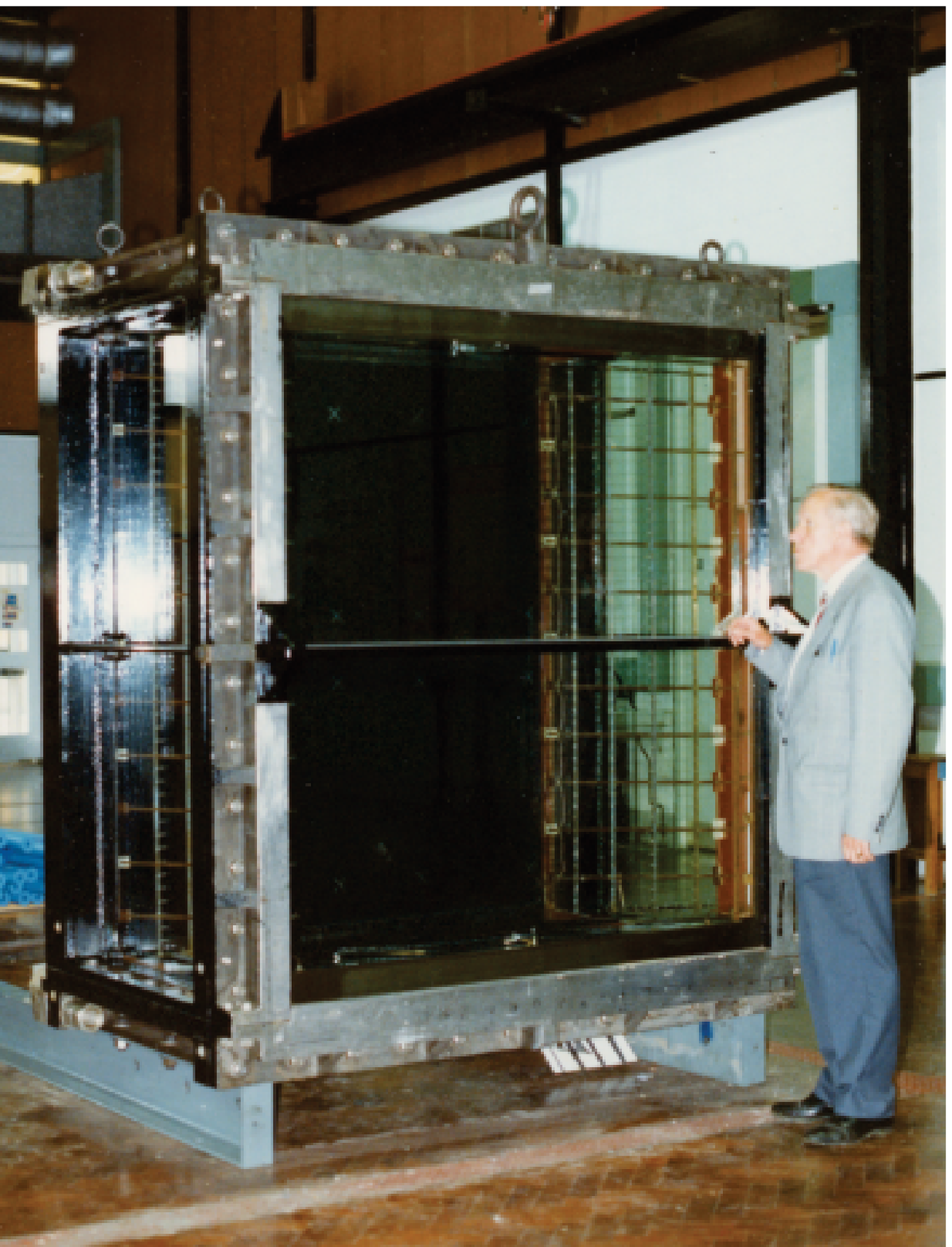}
\end{center}

\pagebreak
\nocite{*}
\bibliographystyle{unsrt}
\bibliography{hodsonbib}

\begin{thebibliography}{10}

\bibitem{Ho48a}
A.L. Hodson.
\newblock Reduction of `insensitive time' in {Geiger-M\"{u}ller} counters.
\newblock {\em Journal of Scientific Instruments}, 25(1):11--13, 1948.

\bibitem{Ho49a}
A.~L. Hodson and A.~Loria.
\newblock Control of a {Wilson} cloud chamber by means of an internal counter.
\newblock {\em Il Nuovo Cimento}, 6:369--370, 1949.

\bibitem{Ho50a}
A.~L. Hodson, A.~Loria, and N.~V. Ryder.
\newblock The control of a {Wilson} cloud chamber by means of an internal
  counter.
\newblock {\em Philosophical Magazine}, 41(319):826 -- 837, 1950.

\bibitem{Ho51a}
A.~L. Hodson.
\newblock The temperature effect of extensive air showers.
\newblock {\em Proceedings of the Physical Society. Section A},
  {64}({384}):{1061--1064}, {1951}.

\bibitem{Ho52a}
A.~L. Hodson.
\newblock {The altitude variation of penetrating showers}.
\newblock {\em Proceedings of the Physical Society. Section A},
  {65}({393}):{702--708}, {1952}.

\bibitem{Ho53a}
A.~L. Hodson.
\newblock {Some aspects of the altitude variation of extensive air showers}.
\newblock {\em Proceedings of the Physical Society. Section A},
  {66}({397}):{49--64}, {1953}.

\bibitem{Ho53b}
A.~L. Hodson.
\newblock Penetrating particles in extensive air showers.
\newblock {\em Proceedings of the Physical Society. Section A}, 66(1):65--72,
  1953.

\bibitem{Ho53c}
J.~Ballam, D.R. Harris, A.L. Hodson, R.R. Rau, G.T. Reynolds, and M.~Vidale.
\newblock {Magnetic field cloud chamber for studying nuclear interactions in
  the cosmic radiation. I. Equipment. (Abstract)}.
\newblock {\em {Physical Review}}, {90}({2}):{369}, {1953}.

\bibitem{Ho53d}
R.R. Rau, M.~Vidale, G.T. Reynolds, D.R. Harris, A.L. Hodson, and J.~Ballam.
\newblock {Magnetic field cloud chamber for studying nuclear interactions in
  the cosmic radiation. II. Preliminary results. (Abstract)}.
\newblock {\em {Physical Review}}, {90}({2}):{369}, {1953}.

\bibitem{Ho53e}
J.~Ballam, D.R. Harris, A.L. Hodson, R.R. Rau, G.T. Reynolds, S.B. Treiman, and
  M.L. Vidale.
\newblock {Momenta of $V^0$ particles. (Abstract)}.
\newblock {\em {Physical Review}}, {91}({2}):{446}, {1953}.

\bibitem{Ho53f}
J.~Ballam, D.R. Harris, A.L. Hodson, R.R. Rau, G.T. Reynolds, S.B. Treiman, and
  M.~Vidale.
\newblock {Kinetic energies of $V_1^{0}$ particles}.
\newblock {\em {Physical Review}}, {91}({4}):{1019--1020}, {1953}.

\bibitem{Ho54a}
D.R. Harris and A.L. Hodson.
\newblock {Production of $V^0$ particles in copper. (Abstract)}.
\newblock {\em {Physical Review}}, {95}({2}):{661}, {1954}.

\bibitem{Ho54b}
W.H. Arnold, J.~Ballam, H.~Gursky, A.L. Hodson, R.R. Rau, G.T. Reynolds, and
  S.B. Treiman.
\newblock {Unstable cosmic-ray particles observed in a double cloud chamber
  arrangement. (Abstract)}.
\newblock {\em {Physical Review}}, {96}({3}):{853}, {1954}.

\bibitem{Ho54c}
A.L. Hodson, J.~Ballam, W.H. Arnold, D.R. Harris, R.R. Rau, G.T. Reynolds, and
  S.B. Treiman.
\newblock {Cloud-chamber evidence for a charged counterpart of the $\theta^0$
  particle}.
\newblock {\em {Physical Review}}, {96}({4}):{1089--1095}, {1954}.

\bibitem{Ho55a}
W.H. Arnold, J.~Ballam, A.L. Hodson, R.R. Rau, G.T. Reynolds, S.B. Treiman, and
  V.A. {Van~Lint}.
\newblock {Analysis of charged $V$ events. (Abstract)}.
\newblock {\em {Physical Review}}, {98}({1}):{275--276}, {1955}.

\bibitem{Ho55b}
J.~Ballam, A.L. Hodson, and G.T. Reynolds.
\newblock {Observations on $S$ particles}.
\newblock {\em {Physical Review}}, {99}({3}):{1038}, {1955}.

\bibitem{Ho55c}
S.B. Treiman, G.T. Reynolds, and A.L. Hodson.
\newblock {Angular correlation effects in unstable particle decay}.
\newblock {\em {Physical Review}}, {97}({1}):{244--245}, {1955}.

\bibitem{Ho55d}
J.~Ballam, A.L. Hodson, W.~Martin, R.R. Rau, G.T. Reynolds, and S.B. Treiman.
\newblock {Orientation of planes in double $V^0$ decay events}.
\newblock {\em {Physical Review}}, {97}({1}):{245--246}, {1955}.

\bibitem{Ho56a}
W.H. Arnold, J.~Ballam, A.L. Hodson, G.K. Lindeberg, R.R. Rau, G.T. Reynolds,
  and S.B. Treiman.
\newblock {Recent cloud chamber results on masses of heavy mesons and
  hyperons}.
\newblock {\em {Il Nuovo Cimento}}, {4, Supp.2}:{559--564}, {1956}.

\bibitem{Ho65a}
A.L. {Hodson}, T.C. {Bacon}, and B.R. {Pullan}.
\newblock A large {Wilson} cloud chamber and its application to ionization
  calorimetry work.
\newblock In {\em 9th International Cosmic Ray Conference, London}, volume~2,
  pages 1082--1084, 1965.

\bibitem{Ho65b}
D.A. {Briggs} and A.L. {Hodson}.
\newblock Particle counting in locally produced showers using vidicon
  television techniques.
\newblock In {\em 9th International Cosmic Ray Conference, London}, volume~2,
  pages 1105--1107, 1965.

\bibitem{Ho71a}
W.E. {Hazen}, A.L. {Hodson}, and D.F. {Winterstein}.
\newblock A cosmic-ray quark search with a large cloud chamber.
\newblock In {\em 12th International Cosmic Ray Conference, Hobart}, volume~3,
  page 1161, 1971.

\bibitem{Ho73a}
W.E. {Hazen}, A.L. {Hodson}, D.~{Winterstein}, and O.~{Keller}.
\newblock Search for e/3 quarks in cosmic rays with the {Leeds} cloud chamber.
\newblock In {\em 13th International Cosmic Ray Conference, Denver}, volume~3,
  pages 2087--2089, 1973.

\bibitem{Ho73b}
W.E. {Hazen}, A.L. {Hodson}, D.~{Winterstein}, and O.~{Keller}.
\newblock Transverse momenta from an air-shower study.
\newblock In {\em 13th International Cosmic Ray Conference, Denver}, volume~3,
  pages 2124--2125, 1973.

\bibitem{Ho74a}
D.F. Winterstein, W.~E. Hazen, A.~L. Hodson, B.~Green, and J.~Kass.
\newblock Search for quarks in dense central region of air showers.
  {(Abstract)}.
\newblock {\em Bulletin of the American Physical Society}, 19(4):583, 1974.

\bibitem{Ho74b}
W.~E. Hazen, A.~L. Hodson, B.~Green, J.~Kass, and D.F. Winterstein.
\newblock {Search for Mandelas with Leeds cloud chamber. (Abstract)}.
\newblock {\em Bulletin of the American Physical Society}, 19(4):583, 1974.

\bibitem{Ho74c}
O.~Keller, W.~E. Hazen, D.F. Winterstein, and A.~L. Hodson.
\newblock {Cloud chamber evidence for large $p_\bot$ in cosmic-ray air showers.
  (Abstract)}.
\newblock {\em Bulletin of the American Physical Society}, 19(4):447, 1974.

\bibitem{Ho75a}
W.E. {Hazen}, J.R. {Kass}, B.R. {Green}, and A.L. {Hodson}.
\newblock A cloud chamber study of large cosmic ray bursts from an absorber at
  sea level. ({Abstract}).
\newblock In {\em 14th International Cosmic Ray Conference, M\"{u}nchen},
  volume~7, page 2461, 1975.

\bibitem{Ho75b}
W.E. {Hazen}, A.L. {Hodson}, J.R. {Kass}, B.R. {Green}, and P.G. {Lloyd}.
\newblock A search for e/3 quarks using the {Leeds} cloud chamber.
  ({Abstract}).
\newblock In {\em 14th International Cosmic Ray Conference, M\"{u}nchen},
  volume~7, page 2473, 1975.

\bibitem{Ho75c}
W.E. {Hazen}, B.R. {Green}, A.L. {Hodson}, and J.R. {Kass}.
\newblock A search for precursors to extensive air showers. ({Abstract}).
\newblock In {\em 14th International Cosmic Ray Conference, M\"{u}nchen},
  volume~7, page 2485, 1975.

\bibitem{Ho75d}
W.E. {Hazen}, A.L. {Hodson}, O.A. {Keller}, B.R. {Green}, and J.R. {Kass}.
\newblock Observations of core structure in extensive air showers of
  $10^{15}$--$10^{16}$ {eV}.
\newblock In {\em 14th International Cosmic Ray Conference, M\"{u}nchen},
  volume~8, pages 2984--2988, 1975.

\bibitem{Ho75e}
A.L. {Hodson}.
\newblock Possible suppression of {{\v{C}}erenkov} emission in
  closely-collimated jets. {An} explanation of the `magnetic monopole' event?
\newblock In {\em 14th International Cosmic Ray Conference, M\"{u}nchen},
  volume~12, pages 4065--4067, 1975.

\bibitem{Ho75f}
W.E. Hazen, A.L. Hodson, D.F. Winterstein, B.R. Green, J.R. Kass, and O.A.
  Keller.
\newblock Search for $\frac{1}{3}e$ quarks in cosmic rays with the {Leeds}
  cloud chamber.
\newblock {\em Nuclear Physics B}, 95(2):189--196, 1975.

\bibitem{Ho75g}
W.E. {Hazen}, B.R. {Green}, A.L. {Hodson}, and J.R. {Kass}.
\newblock A search for precursors to extensive air showers.
\newblock {\em Nuclear Physics B}, 96:401--406, 1975.

\bibitem{Ho78a}
A.L. {Hodson}, J.M. {Foster}, B.R. {Green}, and R.M. {Bull}.
\newblock A 35 sq m array of current-limited spark chambers for the
  investigation of extensive air shower cores.
\newblock In {\em 15th International Cosmic Ray Conference, Plovdiv},
  volume~12, pages 160--165, 1978.

\bibitem{Ho78b}
J.M. {Foster}, A.L. {Hodson}, B.R. {Green}, W.E. {Hazen}, A.Z. {Hendel}, and
  R.M. {Bull}.
\newblock {Observations of extensive air shower cores in the energy range
  $10^{14}$--$10^{16}$ {eV}}.
\newblock In {\em 15th International Cosmic Ray Conference, Plovdiv},
  volume~12, pages 50--55, 1978.

\bibitem{Ho79a}
W.~E. Hazen, A.~Z. Hendel, J.~F. Foster, B.~A. Green, A.~L. Hodson, and
  R.~Bull.
\newblock High $p_t$ cross section from multiple core rates in air showers.
\newblock {\em AIP Conference Proceedings}, 49(1):31--39, 1979.

\bibitem{Ho79b}
M.R. {Porter} and A.L. {Hodson}.
\newblock A study of the air-shower response of current-limited spark chambers
  and scintillators. {(Abstract)}.
\newblock In {\em 16th International Cosmic Ray Conference, Kyoto}, volume~8,
  page 227, 1979.

\bibitem{Ho79c}
W.L. {Hazen}, A.Z. {Hendel}, J.M. {Foster}, B.R. {Green}, A.L. {Hodson}, M.R.
  {Porter}, and R.M. {Bull}.
\newblock Air shower subcores in air and high $p_t$.
\newblock In {\em 16th International Cosmic Ray Conference, Kyoto}, volume~8,
  pages 230--235, 1979.

\bibitem{Ho79d}
B.R. {Green}, A.~{Ash}, and A.L. {Hodson}.
\newblock Computer simulations of extensive air shower cores of energy
  $10^{14}$--$10^{16}$ {eV}.
\newblock In {\em 16th International Cosmic Ray Conference, Kyoto}, volume~13,
  pages 217--222, 1979.

\bibitem{Ho79e}
B.R. {Green} and A.L. {Hodson}.
\newblock {A Monte-Carlo study of detection methods for extensive air shower
  cores falling within a given area}.
\newblock In {\em 16th International Cosmic Ray Conference, Kyoto}, volume~13,
  pages 211--216, 1979.

\bibitem{Ho79f}
A.M. {Hillas}, J.M. {Foster}, B.R. {Green}, A.L. {Hodson}, W.E. {Hazen}, and
  A.Z. {Hendel}.
\newblock {EM transition effect for air shower subcores}.
\newblock In {\em 16th International Cosmic Ray Conference, Kyoto}, volume~8,
  pages 236--239, 1979.

\bibitem{Ho81a}
R.M. {Porter}, L.A. {Hodson}, and R.M. {Bull}.
\newblock A study of the air-shower response of current-limited spark chambers.
\newblock In {\em 17th International Cosmic Ray Conference, Paris}, volume~11,
  pages 365--368, 1981.

\bibitem{Ho81b}
W.E. Hazen, A.G. Ash, J.M. Foster, A.L. Hodson, and M.R. Porter.
\newblock Simulation studies of uncertainties in air shower subcore
  determination of $p_t$.
\newblock In {\em 17th International Cosmic Ray Conference, Paris}, volume~5,
  page 148, 1981.

\bibitem{Ho81c}
W.E. Hazen, A.Z. Hendel, A.G. Ash, J.M. Foster, A.L. Hodson, M.R. Porter, and
  R.M. Bull.
\newblock Subcore observations and high $p_t$.
\newblock In {\em 17th International Cosmic Ray Conference, Paris}, volume~5,
  pages 144--147, 1981.

\bibitem{Ho81d}
A.L. Hodson, M.R. Porter, and R.M. Bull.
\newblock A new search for free e/3 quarks from
  $\sim10^{15}$--$\,5\!\cdot\!10^{15}$ {eV} interactions. {(Abstract)}.
\newblock In {\em 17th International Cosmic Ray Conference, Paris}, volume~5,
  page~50, 1981.

\bibitem{Ho81e}
E.W. {Hazen}, Z.A. {Hendel}, G.A. {Ash}, M.J. {Foster}, L.A. {Hodson}, R.M.
  {Porter}, and M.R. {Bull}.
\newblock Measurements of electron densities in {E.A.S.} using spark chambers
  and thin scintillators. {(Abstract)}.
\newblock In {\em 17th International Cosmic Ray Conference, Paris}, volume~6 of
  {\em International Cosmic Ray Conference}, page 176, 1981.

\bibitem{Ho81i}
M.R. {Porter}, J.M. {Foster}, A.L. {Hodson}, W.E. {Hazen}, A.Z. {Hendel}, and
  R.M. {Bull}.
\newblock {Air shower density spectrum}.
\newblock In {\em 17th International Cosmic Ray Conference, Paris}, volume~11,
  pages 417--420, 1981.

\bibitem{Ho81aaa}
W.E.Hazen, A.Z. Hendel, A.G.Ash, J.M. Foster, B.R. Green, A.L. Hodson, M.R.
  Porter, and R.M. Bull.
\newblock Subcores in cosmic-ray air showers and high transverse momentum.
\newblock {\em Journal of Physics G: Nuclear Physics}, 7(9):1285--1296, 1981.

\bibitem{Ho82b}
W.E. {Hazen}, A.Z. {Hendel}, J.M. {Ash}, A.G.~{Foster}, A.L. {Hodson}, M.R.
  {Porter}, and R.M. {Bull}.
\newblock Test of high $p_t$ scaling from cosmic-ray interactions up to 400
  {TeV}.
\newblock {\em Physical Review Letters}, 48:1562--1564, 1982.

\bibitem{Ho83a}
R.M. {Bull}, A.L. {Hodson}, and M.R. {Porter}.
\newblock A recording system for scintillation counter pulses over a wide
  dynamic range.
\newblock In {\em 18th International Cosmic Ray Conference, Bangalore},
  volume~9, pages 456--459, 1983.

\bibitem{Ho83b}
A.G. {Ash} and A.L. {Hodson}.
\newblock Comparison of simulation results with experimental data on
  $10^{14}$--$10^{16}$ {eV} air showers.
\newblock In {\em 18th International Cosmic Ray Conference, Bangalore},
  volume~11, pages 205--208, 1983.

\bibitem{Ho83c}
A.L. {Hodson}, M.R. {Porter}, A.G. {Ash}, and R.M. {Bull}.
\newblock Core flattening in $10^{14}$--$10^{16}$ {eV} air showers.
\newblock In {\em 18th International Cosmic Ray Conference, Bangalore},
  volume~11, pages 201--204, 1983.

\bibitem{Ho85a}
A.L. {Hodson}, A.G. {Ash}, and R.M. {Bull}.
\newblock {Particle distributions in $\sim 10^{14}$--$10^{16}$ {eV} air shower
  cores at sea level}.
\newblock In {\em 19th International Cosmic Ray Conference, La Jolla},
  volume~7, pages 81--84, 1985.

\bibitem{Ho85b}
A.L. {Hodson}, R.M. {Bull}, R.S. {Taylor}, and C.H. {Belford}.
\newblock Progress report on a new search for free e/3 quarks in the cores of
  $10^{14}$--$10^{16}$ {eV} air showers.
\newblock In {\em 19th International Cosmic Ray Conference, La Jolla},
  volume~8, pages 306--309, 1985.

\bibitem{Ho87a}
P.D. {Acton}, A.G. {Ash}, R.M. {Bull}, and A.L. {Hodson}.
\newblock The hadron component within a few metres of the axes of
  $10^{14}$--$10^{16}$ {eV} air showers.
\newblock In {\em 20th International Cosmic Ray Conference, Moscow}, volume~6,
  pages 51--53, 1987.

\bibitem{Ho87b}
A.G. {Ash}, R.M. {Bull}, A.L. {Hodson}, and R.S. {Taylor}.
\newblock A trigger circuit for selecting air shower cores of
  $10^{14}$--$10^{16}$ {eV} primary energy, using particle-density ratios.
\newblock In {\em 20th International Cosmic Ray Conference, Moscow}, volume~6,
  pages 462--465, 1987.

\bibitem{Ho87c}
C.H. {Belford}, R.M. {Bull}, A.L. {Hodson}, and R.S. {Taylor}.
\newblock Quark limits in air shower cores.
\newblock In {\em 20th International Cosmic Ray Conference, Moscow}, volume~6,
  pages 352--355, 1987.

\bibitem{Ho89a}
P.~D. Acton, A.~G. Ash, A.~L. Hodson, and R.~M. Bull.
\newblock The hadron component of $10^{14}$--$10^{16}$ {eV} extensive air
  showers.
\newblock In {\em 21st International Cosmic Ray Conference, Adelaide},
  volume~9, pages 264--267, 1989.

\bibitem{Ho91a}
A.L. {Hodson}.
\newblock {Reminiscences on experimental work with various particle detectors}.
\newblock {\em Nuclear Physics B Proceedings Supplements}, 22:178--196, 1991.

\end{thebibliography}

\end{document}